\documentclass[11pt]{article}
\usepackage{moriond,epsfig,amsmath}

\def\Journal#1#2#3#4{{#1} {\bf #2}, #3 (#4)}

\def\NPB{{\em Nucl. Phys.} B}
\def\PLB{{\em Phys. Lett.} B}
\def\PRL{\em Phys. Rev. Lett.}
\def\PRD{{\em Phys. Rev.} D}

\def\JHP{\em JHEP}
\def\EJC{{\em Eur. Phys. J.} C}

\def\be{\begin{equation}}
\def\ee{\end{equation}}
\def\bea{\begin{eqnarray}}
\def\eea{\end{eqnarray}}
\begin{document}
\vspace*{4cm}
\title{SUPER-LEADING LOGARITHMS IN GAPS-BETWEEN-JETS}

\author{A. KYRIELEIS}

\address{School of Physics \& Astronomy, The University of Manchester,\\
Manchester M13 9PL, U.K.}

\maketitle
\abstracts{We identify a source of super-leading logarithms in
  the  gaps-between-jets observable at hadron colliders. These new
  contributions are expected to generally appear in  non-global
  observables in QCD and  are connected with the presence of Coulomb
  phase terms.}

\section{Introduction}
The resummation of large logarithms associated with wide angle soft gluon
emissions has been investigated for the last 20 years. 
For certain observables the  contributions from non-global
logarithms \cite{NGL}  have to be taken into account.
One of the simplest of these non-global observables is the
`gaps-between-jets' cross-section. This is the cross-section for
producing a pair of high transverse momentum jets (Q) with a restriction
on the transverse momentum of any additional jets radiated in between
the two jets, i.e. $k_T < Q_0$ for emissions in the gap. This
observable has been studied \cite{OdSter,AppSey} and has been measured
at HERA and the Tevatron \cite{gaps exp}. 

\begin{figure}[h]
\begin{center}
\includegraphics[height=2.5in]{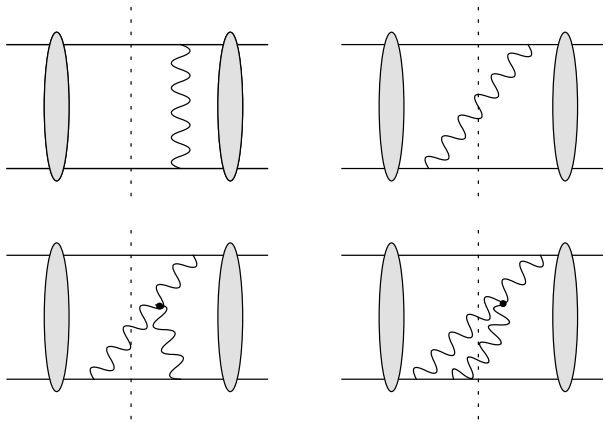}\\[0pt]
\end{center}
\caption{Illustrating the cancellation (and miscancellation) of soft gluon
corrections.}%
\label{fig:miscancel}%
\end{figure}

In the original calculations \cite{OdSter} of the gaps-between-jets
cross section,  all those terms $\sim\alpha_{s}^{n}\ln^{n}(Q/Q_{0})$  that can be obtained by dressing the primary
$2\rightarrow2$ scattering in all possible ways with soft virtual gluons were
summed. 
The restriction to soft gluons implies the use of the
eikonal approximation. Let us  focus on   quark-quark scattering from
now on. The corresponding resummed cross-section can be written
\begin{equation}
\sigma=\mathbf{M}^{\dag}\mathbf{S}_{V}\mathbf{M}\qquad\mbox{with}\qquad
\mathbf{M}=\exp\left(  -\frac{2\alpha_{s}}{\pi}%
{\displaystyle\int\limits_{Q_{0}}^{Q}}
\frac{dk_{T}}{k_{T}}\mathbf{~\Gamma}\right)  \mathbf{M}_0.\label{eq:OS}%
\end{equation}
Here, $\mathbf{M}$ is the all-orders $qq\rightarrow qq$  amplitude (a 2-component vector in colour space), $\mathbf{M}_0$ is the hard scattering
amplitude and $\mathbf{S}_V$ represents the cut.
The anomalous dimension matrix
$\Gamma$ \cite{Gamma} incorporates the effect of dressing a
$qq\rightarrow qq$
amplitude with a  virtual gluon in all possible ways.
It receives contributions from  two distinct regions of the
loop-integral: the first corresponds to an on-shell gluon (to which
one can  assign a rapidity) and is
identical, but with  opposite sign, to the contribution from a
real gluon.
The second contribution, sometimes referred to as the `Coulomb gluon
\cite{Coulomb} contribution'  is
purely imaginary ($i\pi$ terms)  and stems from the region where the emitting parton is
on-shell.
 Eq.(\ref{eq:OS}) therefore corresponds to the  independent
 emission of soft gluons, i.e. the iterative dressing of the
 $2\rightarrow 2$ process with a softer gluon:  due to perfect real/virtual
 cancellation  outside the gap (the first line of fig.\ref{fig:miscancel}
 shows two contributions) one only has to consider virtual gluons
 in the gap and the Coulomb terms.

However, there is another source of leading logarithms. Let us
consider the two diagrams in the second line of
fig.\ref{fig:miscancel}. A real gluon (which is outside the gap by the
definition of our observable) emits a softer real or virtual
gluon. The real-virtual cancellation is guaranteed only for the softest
gluon. Since real gluons above $Q_0$ are forbidden in the gap, the
two diagrams do not completely cancel; the left diagram with the virtual gluon
being in the gap and its $k_T$ being larger than $Q_0$ survives. The
non-global nature of our observable has prevented the soft gluon
cancellation which is necessary in order that eq.(\ref{eq:OS}) should be
the complete result. 

It is therefore necessary to include the emission of any
number of soft gluons outside the gap region (real and virtual)
dressed with any number of virtual gluons within the gap
region. Clearly it is a formidable challenge to sum all leading
logarithms, mainly because of the complicated colour
structure. Progress has been made, working in the large N
approximation \cite{AppSey}. Here, we keep the exact colour structure
but instead we only compute the cross-section for one gluon outside
the gap region. This can be viewed as the first term in an expansion
in the number of out-of-gap-gluons.
\section{Super-leading logarithms}
In order to extract the leading logarithms we consider soft gluons
strongly ordered in transverse momentum.
The cross-section for one gluon outside and any number
of gluons inside the gap  is split into two parts corresponding to a
virtual or real  out-of-gap gluon: 
\begin{align}
&\sigma_1    =-\bar\alpha\int_{Q_{0}}^{Q}\frac{dk_{T}}{k_{T}%
}~\int\limits_{\text{out}}\frac{dy~d\phi}{2\pi}\:\left(A_V+
  A_R\right)\:,\qquad\bar\alpha\equiv \frac{2\alpha_s}{\pi}
\label{eq:sig1}
\end{align}
\begin{align}
 A_R= \mathbf{M}_{0}^{\dag}\exp\left(  {\small -}\bar\alpha%
\int\limits_{k_{T}}^{Q}\frac{dk_{T}^{\prime}}{k_{T}^{\prime}}\mathbf{\Gamma
}^{\dag}\right)  &\mathbf{D}_{\mu}^{\dag}\exp\left(  {\small -}\bar\alpha\int\limits_{Q_{0}}^{k_{T}}\frac{dk_{T}^{\prime}}%
{k_{T}^{\prime}}\mathbf{\Lambda}^{\dag}\right)  \mathbf{S}_{R}
\nonumber\\
 \exp\left(  {\small -}\bar\alpha\int\limits_{Q_{0}}^{k_{T}%
}\frac{dk_{T}^{\prime}}{k_{T}^{\prime}}\mathbf{\Lambda}\right)  \mathbf{D}&
^{\mu}\exp\left(  {\small -}\bar\alpha\int\limits_{k_{T}}%
^{Q}\frac{dk_{T}^{\prime}}{k_{T}^{\prime}}\mathbf{\Gamma}\right)
\mathbf{M}_{0}\;, \\ 
A_V=   \mathbf{M}_{0}^{\dag}\exp\left({\small -}\bar \alpha \int\limits_{Q_{0}}^{Q}\frac{dk_{T}^{\prime}}{k_{T}^{\prime}}\mathbf{\Gamma
}^{\dag}\right)  \mathbf{S}_{V} \exp&\left({\small -}\bar \alpha
\int\limits_{Q_{0}}^{k_{T}}\frac{dk_{T}^{\prime}}{k_{T}^{\prime}}\mathbf{\Gamma}\right)
~\boldsymbol{\gamma}\mathbf{~}\exp\left({\small -}\bar\alpha\int\limits_{k_{T}}^{Q}\frac{dk_{T}^{\prime}}{k_{T}^{\prime}}\mathbf{\Gamma
}\right)  \mathbf{M}_{0}~{\small +~}\text{{\small c.c.}}.
\label{eq:virtual}%
\end{align}
$\mathbf{D}^\mu$ and $\boldsymbol{\gamma}$ are the matrices that  
represent the emission of a real and a virtual gluon ($k_T, y, \phi$)
outside the gap, respectively.  The major new ingredient is the matrix
$\mathbf{\Lambda}$ \cite{LAM} which  incorporates the dressing
of the $qq\to qqg$ process with a virtual gluon. The emission of the
out-of-gap gluon is sandwiched between two exponentials: this accounts
for all possible positions of the out-of-gap gluon within a chain of
any number of $k_T$-ordered gluons within the gap. 

The phase space of the out-of-gap gluon in eq.(\ref{eq:sig1})  includes the
configurations where it is collinear to either of the external
quarks. One might suppose that the corresponding divergences cancel
among $A_R$ and $A_V$. This is true in case of the final state
collinear limit. However, in the limit of the out-of-gap gluon
becoming collinear to one of the initial state quarks,
i.e. $|y|\to\infty, k_T>Q_0$, there is no cancellation:
\begin{equation}
[A_V+A_R]_{|y|\to\infty} \ne 0.
\label{eq:oneout}
\end{equation}
In particular, $(A_V+A_R)$ becomes independent of $y$ in that limit.
This has severe consequences. As the out-of-gap region stretches to
infinity in  rapidity, the integral eq.(\ref{eq:sig1}) is divergent as it
stands. This divergence however indicates that one needs to go beyond
the soft approximation when considering the out-of-gap gluon. 
 As the  energy of the out-of-gap gluon is
$E_g=1/2\, k_T(e^y+e^{-y})$ and $k_T>Q_0$ the limit $|y|\to \infty$
implies $E_g\to \infty$. Insisting on $E_g$ being smaller than the
centre-of-mass energy $E_{CMS}=2 Q \cosh \Delta/2$ ($\Delta$ is the
rapidity interval between the final state quarks) then translates into 
 $y<\ln(Q/k_T)$. The integration in eq.(\ref{eq:sig1}) therefore gives 
 $\alpha_s\ln^2(Q/Q_0)$ whereas each gluon within the gap only
 contributes  $\alpha_s\ln(Q/Q_0)$. We have obtained a hitherto unknown
 super-leading
 logarithmic contribution, i.e. a term  $\sim \alpha_s^n\ln^{n+1}(Q/Q_0)$.

In a more rigorous treatment \cite{SLL} the out-of-gap gluon is
considered in the collinear (but not soft) approximation. The `plus
prescription' for the collinear divergence appears accompanied by an
additional term that generates the super-leading logarithm. The
additional logarithm can therefore be seen as extra
collinear logarithm due to the failure of the `plus prescription'
for gluons with $k_T > Q_0$; collinear logarithms
in the gaps-between-jets cross-section can be summed into the pdf's only up to scale $Q_0$.

The super-leading logarithms are formally more important than any
leading-logarithmic result. Their numerical impact, though, still has to be
investigated (for first results see \cite{SLL}). There is a  couple of
remarks to be made.

\begin{itemize}
\item 
The miscancellation eq.(\ref{eq:oneout}) and hence the super-leading
logarithm is intimately connected with
the Coulomb phase terms. If one artificially switches off the $i \pi$
terms in the evolution matrices, then there is full cancellation  in
eq.(\ref{eq:oneout}). Moreover, the super-leading logarithm makes its
appearance at the lowest possible order in $\alpha_s$, i.e. at 
$O(\alpha_s^4)$ relative to the Born cross-section. More explicitly, the 
$O(\alpha_{s})$ and $O(\alpha_{s}^{2})$ corrections to the Born cross-section simply never involve
more than one $i\pi$ term and hence any $i\pi$ terms must cancel since the
cross-section is real. The first candidate order at which two factors of
$i\pi$ can appear is therefore $O(\alpha_{s}^{3})$. However, the addition of
the gluon with the lowest $k_{T}$ can never generate a net factor of $i\pi$
since any such factors must cancel between the two diagrams where the lowest
$k_{T}$ gluon lies either side of the cut. The first $i\pi$ terms and
the first 
super-leading logarithm appear in case of four  soft gluons:%
\begin{equation}
\sigma_1\sim\sigma_{0}\left(  \frac{2\alpha_{s}}{\pi}\right)  ^{4}\ln^{5}\left(
\frac{Q}{Q_{0}}\right)  \pi^{2}Y.\label{eq:sllform}%
\end{equation}
Here, $Y$ is the size of the rapidity gap and $\sigma_0$ is the Born
cross-section.
\item
At $O(\alpha_s^4)$  the super-leading logarithms arise from   one
out-of-gap gluon  at maximum. The proof of this is
based on the fact that two or more out-of-gap
gluons imply two or less gluons that can provide $i\pi$ terms. We have
shown above that this is not enough for the $i\pi$ factors
to appear in the cross-section and we know that only if 
$i\pi$ terms survive  there is a non-cancellation in the initial
state collinear limit. 
\item 
At higher orders in $\alpha_s$ more gluons can be outside the gap. At
each order there is a maximum number of out-of-gap gluons that we
expect to provide
the maximum power of the logarithm stemming from the region where they
all  become collinear to either of the initial state partons. 
At each higher order we then have an additional
$\alpha_s\ln^2(Q/Q_0)$ in such a configuration. What appears as  super-leading
logarithms as compared to the single logs, $\alpha^n_s\ln(Q/O_0)^n$
would thus constitute a series of  double logarithms,
$\alpha^n_s(\ln^2(Q/O_0))^n$. To resum these double
logarithms a deeper understanding of the colour evolution of
multi-parton systems seems necessary.
\item 
The new super-leading contributions are not restricted to the
gaps-between-jets observable. In event shape variables, for instance,
the requirement of the variable to be smaller than some value translates
into a restriction of the phase space available for real gluons, i.e. a
gap. We expect the super-leading logarithms  to  arise generally
in non-global observables.
\end{itemize}
%
%
\section*{Acknowledgements}
The presented work was done in collaboration with M.~Seymour and
J.~Forshaw. I like to thank  L.~Motyka and G.~Salam for very fruitful
discussions during the Moriond meeting.

\newpage
\end{document}